\documentclass[prl,twocolumn,nofootinbib, preprintnumbers, superscriptaddress]{revtex4}

\usepackage{xr-hyper}

\makeatletter
\newcommand*{\addFileDependency}[1]{
  \typeout{(#1)}
  \@addtofilelist{#1}
  \IfFileExists{#1}{}{\typeout{No file #1.}}
}
\makeatother
 
\newcommand*{\myexternaldocument}[1]{%
    \externaldocument{#1}%
    \addFileDependency{#1.tex}%
    \addFileDependency{#1.aux}%
}

\myexternaldocument{supplemental-material}

\usepackage{amsmath,amssymb,slashed,braket}
\usepackage{graphicx}
\usepackage{epstopdf}
\usepackage{float,appendix}
\usepackage[colorlinks=true,
            linkcolor=blue,
            urlcolor=blue,
            citecolor=green,          
            bookmarks=true,
            bookmarksnumbered=true,
            breaklinks=true,
            pdfpagemode=Fullscreen,
            pdfstartview=FitBH]{hyperref}

\usepackage{esint}

\usepackage[normalem]{ulem}
\usepackage{color}

\definecolor{Orange}{cmyk}{0,0.61,0.87,0}
\definecolor{JungleGreen}{cmyk}{0.99,0,0.52,0}
\definecolor{OliveGreen}{cmyk}{0.64,0,0.95,0.40}
\definecolor{Brown}{cmyk}{0,0.81,1,0.60}
\definecolor{RoyalBlue}{cmyk}{0.71,0.53,0,0.12}
\definecolor{Gray}{cmyk}{0,0,0,0.40}
\definecolor{LightPink}{cmyk}{0.0,0.25,0,0}
\definecolor{LLightPink}{cmyk}{0.0,0.10,0,0}
\definecolor{LightBlue}{cmyk}{0.25,0,0,0}
\definecolor{LightGray}{cmyk}{0,0,0,0.2}

\usepackage{xcolor}
\definecolor{gesfpurple}{rgb}{0.47,0.19,0.42}

\definecolor{gesflanse}{rgb}{0.00,0.50,0.50}

\definecolor{gesfblue}{rgb}{0.08,0.42,0.76}

\definecolor{gesfred}{rgb}{1,0,0}

\definecolor{gesfwhite}{rgb}{1,1,1}

\definecolor{gesfblack}{rgb}{0,0,0}

\newcommand{\geqn}[1]{Eq.\,\hypersetup{linkcolor=blue}(\ref{#1})\hypersetup{linkcolor=blue}}
\newcommand{\gfig}[1]{{\hypersetup{linkcolor=violet}Fig.\,\ref{#1}\hypersetup{linkcolor=blue}}}

\graphicspath{{figs/}}

\begin{document}

\title{Background-Enhanced Axion Force by Axion Dark Matter}

\author{Yu Cheng}
\email{chengyu@kaist.ac.kr}
\affiliation{Department of Physics, Korea Advanced Institute of Science and Technology (KAIST), Daejeon 34141, South Korea}

\author{Shuailiang Ge}
\email{shuailiangge@kaist.ac.kr}
\affiliation{Department of Physics, Korea Advanced Institute of Science and Technology (KAIST), Daejeon 34141, South Korea}

\begin{abstract}
We investigate the influence of axion dark matter as a background on the spin-independent axion forces between nucleons. Notably, we find that the potential for axion forces scales from $1/r^3$ in a vacuum-only context to $1/r$ when the background effect is considered. Also, the magnitude of the axion force is substantially amplified in proportion to the number density of axion DM particles. These enhancements significantly improve the constraints on the axion decay constant by several orders of magnitude, across a broad range of axion masses, based on the fifth-force experiments such as the Casimir-less and torsion balance tests. This suggests that such experiments are more effective than previously understood in detecting axions. 
\end{abstract}

\maketitle 

\noindent
{\bf Introduction}--
Since its proposal as a solution to the Strong CP problem~\cite{peccei1977cp, peccei1977constraints, weinberg1978new, wilczek1978problem, kim1979weak, shifman1980can, dine1981simple, Zhitnitsky:1980tq}, the axion has drawn considerable interest from both theoretical and experimental sides. In addition, axions also arise in string theory~\cite{Svrcek:2006yi}. One typical observable signature is that axions can mediate new forces between ordinary matter, making fifth-force searches a potential probe for axions. More specifically, axions are considered to mediate spin-dependent forces~\cite{Haber:1987nx,Costantino:2019ixl,Cong:2024qly} due to their pseudoscalar nature. However, such forces vanish at the tree level between unpolarized macroscopic objects. The leading non-vanishing spin-independent force mediated by axions arises from the exchange of an axion pair at one loop~\cite{Grifols:1994zz,Ferrer:1998ue}. Detailed calculations in previous studies \cite{Grifols:1994zz,Ferrer:1998ue} have shown that in the massless limit for axion when the shift symmetry is respected, the potential scales as $1/r^5$. Recently, it has been pointed out \cite{Bauer:2023czj} that due to the breakdown of axion shift symmetry, the scaling of the potential is modified to be $1/r^3$, which greatly increases the sensitivity of searching for axion forces.

On the other hand, it is known that axions can serve as good dark matter (DM) candidate~\cite{Preskill:1982cy, Abbott:1982af, Dine:1982ah}. This motivates us to investigate how the axion-mediated forces are affected in the presence of axion DM as background. A similar effect has been studied in the context of neutrino forces in the presence of cosmic neutrino background~\cite{Horowitz:1993kw,Ferrer:1998ju,Ferrer:1999ad}, and has recently been extended to various neutrino backgrounds~\cite{Ghosh:2022nzo,Arvanitaki:2022oby,Arvanitaki:2023fij,Ghosh:2024qai} as well as dark forces induced by the exchanging light dark sector particles~\cite{Barbosa:2024pkl,VanTilburg:2024xib}.
An inspiring interpretation of the background-related force was introduced in \cite{VanTilburg:2024xib}, termed as the ``wake force", offering insights from both classical and quantum perspectives. 
The presence of a background field drastically impacts 
processes at the loop level rather than the tree level. For 
instance, the anomalous magnetic moment of Standard Model (SM)
particles receives a large correction from a DM background 
field\footnote{ 
A classical description is proposed in 
\cite{Zhou:2025wax}, drawing an analogy to the ponderomotive
force in optics. The author argues that the effect becomes significant only when the DM mass exceeds the characteristic frequency $\omega_0$ of the $g-2$ experimental.}~\cite{Evans:2023uxh,Arza:2023wou,Evans:2024dty}. This is basically because when doing a loop integral, the 
number density of background particles enters, which can 
dominate over the vacuum counterpart. Consequently, axion-mediated forces are especially vulnerable to the background 
correction, since there is no tree-level diagrams involved.
See also other types of DM background effects~\cite{Fukuda:2021drn, Day:2023mkb, Li:2024bbe, Du:2024tin, Yin:2023jjj, Alonso-Alvarez:2019ssa}.


To reiterate, when the number density of background particles becomes sufficiently large, zero-temperature field theory is no longer valid. 
The axion DM mass is in general very light, ranging from $10^{-22}\,$eV to $O(1)\,$eV (see e.g., Refs.~\cite{Marsh:2015xka, Lin:2019uvt}), implying that the number density of axion DM is enormously large with the DM energy density known to be $\sim 0.4{\rm~GeV}/{\rm cm}^3$. The axion propagator must be modified with such a background present. In the vacuum-to-vacuum case, the exception value of the normal ordered product of annihilation and creation operators is a delta function, $\langle 0| a_{\boldsymbol{p} } a^\dagger_{\boldsymbol{k}} | 0 \rangle = (2 \pi)^3 \delta^3(\boldsymbol{p} - \boldsymbol{k})$. However, for an axion emitted from a background axion ``sea" $|n \rangle$, propagating a distance, and then reabsorbed into the background, the expectation becomes
\begin{equation}
    \langle n| a_{\boldsymbol{p}} a^\dagger_{\boldsymbol{k}} | n \rangle 
    =
    (2 \pi)^3 \delta^3(\boldsymbol{p} - \boldsymbol{k}) n(\boldsymbol{k}).
\end{equation}
$n(\boldsymbol{k})$ denotes the phase space distribution of the axion DM background; integrating $n(\boldsymbol{k})$ over the phase gives its number density. This modifies the axion propagator to be 
\begin{equation}
D(k) = 
\frac{i}{k^2 - m_a^2 + i \epsilon} + 2 \pi n(\boldsymbol{k})
\delta(k^2 - m_a^2),
\label{eq:thermalOP}
\end{equation}
which is the same as the scalar boson propagator in the real-time formalism of thermal field theory. In this work, however, we do not restrict $n(\boldsymbol{k})$ to be Bose-Einstein distribution but instead allow for a more general form that describes the distribution of axion DM.

In this paper, we show that the background effect changes the scaling of the potential for the axion force from $1/r^3$ to $1/r$. Also, the magnitude of the potential is greatly enhanced, which becomes proportional to the number density of axion DM particles. Consequently, for a wide range of axion mass, the constraints on the axion parameter $c_{GG}/f$ is improved by several orders of magnitude based on the fifth-force experiments such as the Casimir-less and torsion balance tests. 

\noindent
{\bf Axion force enhanced in axion DM background}--
In this part, we introduce the general formalism to calculate axions force in the presence of axion DM background. The Lagrangian describing couplings of axions to nucleons, expanded to the linear order of the axion decay constant $f^{-1}$, is given by~\cite{Bauer:2021mvw,Bauer:2023czj,Bauer:2024hfv}
\begin{equation}
    \mathcal{L}^{(1)}
    =
    \frac{1}{2} (\partial a)^2 - \frac{1}{2} m_a^2 a^2 + g_{N} \frac{\partial_\mu a}{2 f} \bar N \gamma^\mu \gamma_5 N.
    \label{eq:shiftinvariantO}
\end{equation}
$N \equiv p,n$ represents either proton or neutron, each with a different coupling to axion,
\begin{align}
     g_{p,n} &= g_0 (c_u + c_d + 2 c_{GG}) \nonumber\\
     &\pm g_A \frac{1}{1 - \tau_a^2} 
    \left( c_u - c_d + 2 c_{GG} \frac{m_d - m_u}{m_d + m_u}\right). 
\end{align}
Here, $\tau \equiv  m_a^2/m_\pi^2$, the parameters $c_{u,d}$ and $c_{GG}$ are the coupling of axion with quarks and gluon in UV theory with coefficients $g_0 = 0.440(44)$ and $g_A = 1.254(16)(30)$~\cite{Liang:2018pis,Bauer:2021mvw}.
\geqn{eq:shiftinvariantO} preserves the axion shift symmetry except for the axion mass term. The shift-symmetry breaking effect will be reflected in higher orders. 
At the quadratic order of $f^{-2}$, the shift symmetry breaking axion-nucleon interactions can be described by the dimensional six operators~\cite{Gasser:1987rb,Bauer:2023czj,Bauer:2024hfv}
\begin{equation}
    \mathcal{L}^{(2)} = c_N \frac{a^2}{f^2} \bar N N,
    \label{eq:ShiftBreakO}
\end{equation}
with 
\begin{equation}\label{eq:shift_symmetry_breaking_term}
    c_N \equiv - c_1 \frac{m_\pi^2 }{2} \frac{4 c^2_{GG} (1 - \tau_a)^2 + (c_u- c_d)^2 \tau_a^2}{(1-\tau_a)^2}
\end{equation}
where  $c_1 = -1.26(14)\,$GeV$^{-1}$~\cite{Alarcon:2012kn}. For axion mass $m_a \ll m_\pi$, $\tau_a \approx 0$ and the quadratic coupling reduces to $c_N = - 2 c_1 m^2_\pi c^2_{GG}$.


For the convenience of later calculations, we rewrite Eq.~\eqref{eq:shiftinvariantO} in the non-derivative basis by making a chiral rotation $N\rightarrow N{\rm e}^{ i g_N \frac{a(x)}{2f} \gamma_5}$, which then can be Taylor expanded in terms of $a/f$.
Collecting all axion-nucleon couplings up to the quadratic order, we have
\begin{align}
\mathcal{L}_{int}
    = 
    - i m_N g_{N} \frac{a}{f} \bar{N} \gamma_5 N 
    + 
    m_N g_{N}^2 \frac{a^2}{2 f^2} \bar N N
    + c_N \frac{a^2}{f^2} \bar{N} N.
    \label{eq:LagrangianND}
\end{align}

\begin{figure}[!t]
    \centering
    \includegraphics[width=0.486
    \textwidth]{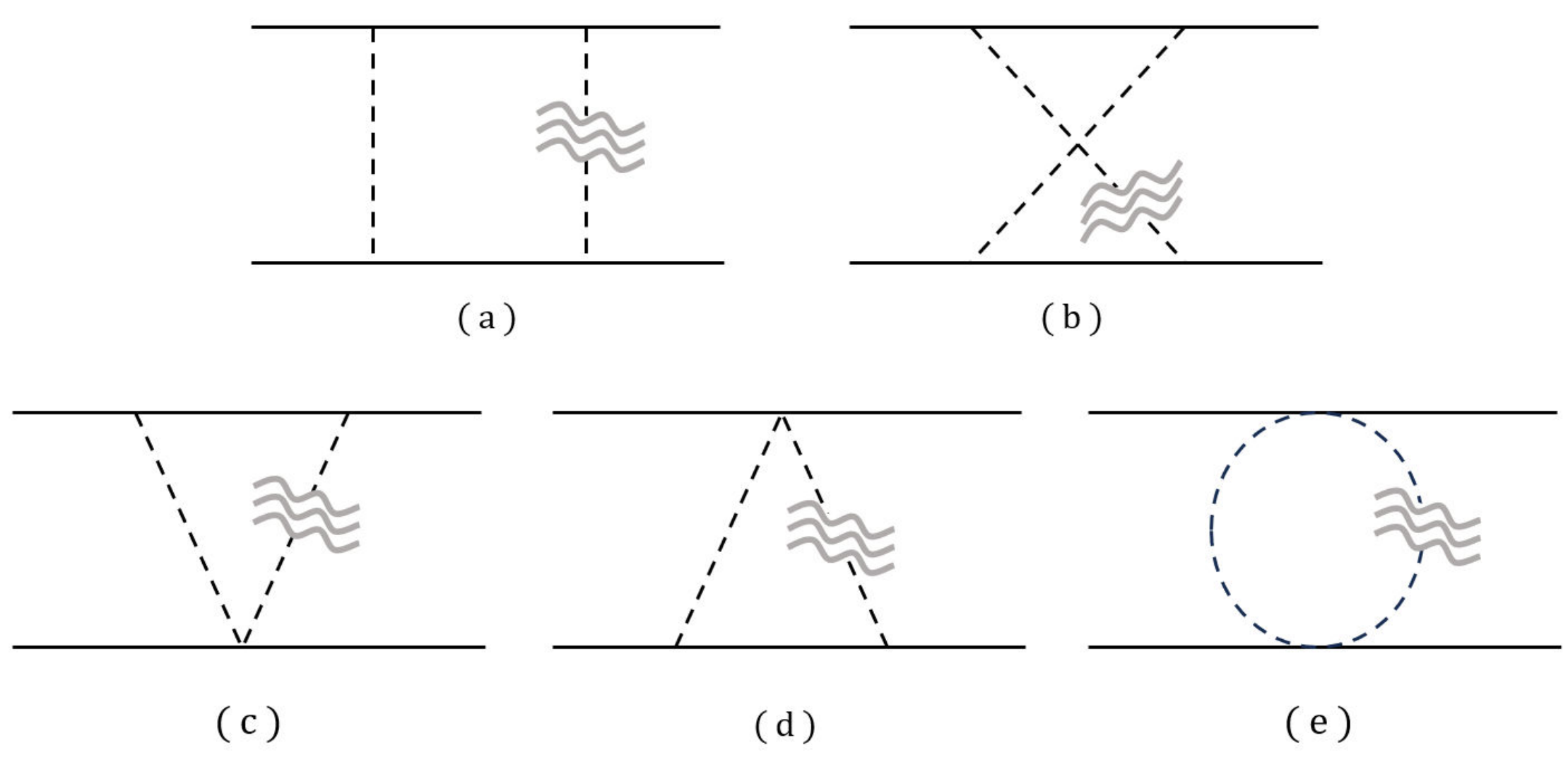}
    \caption{Feynman diagrams which contribute to the potential by exchanging an axion pair. The wavy lines represent the axion DM background. The background-enhancement effect can be obtained by substituting one of the axion propagators with the propagator in Eq.~\eqref{eq:thermalOP}. 
    }
    \label{fig:FeynmanD}
\end{figure}
%


Next, we briefly summarize the steps to calculate the spin-independent axion force between two fermions $N_1$ and $N_2$ by exchanging an axion pair. The force can be viewed as one fermion scattered off the static potential generated by another fermion. According to Born approximation, the potential in phase space, $\tilde{V}(\boldsymbol{q})$, is related to the scattering amplitude via $i \mathcal{M} = - i \tilde{V} (\boldsymbol{q}) 2 m_1 \delta_{s s^\prime} 2 m_2 \delta_{r r^\prime}$ in the non-relativistic limit, where $\boldsymbol{q}$ denotes the momentum transfer.
Then, the potential in position space can be obtained by the Fourier transformation,
\begin{equation}
    V( \boldsymbol{r})
    =
     \int \frac{d^3 \boldsymbol{q}}{(2 \pi)^3} e^{i \boldsymbol{q} \cdot r} \tilde{V}(\boldsymbol{q})
\end{equation}

All five types of diagrams depicting the exchange of an axion pair between two fermions $N_1$ and $N_2$, which contribute to the axion force, are listed in \gfig{fig:FeynmanD}. The three-point
vertex represents the linear coupling $-i m_N g_N /f$ as shown in the first term of 
\geqn{eq:LagrangianND}. There are two types of four-point vertices, $m_N g_N^2/2f^2$ and $c_N g_N^2/f^2$, corresponding to the second and the third terms in \geqn{eq:LagrangianND}, respectively. We 
call the second term the shift-symmetry-invariant (SI) operator while the third term the shift-symmetry-breaking (SB) operator. One should not be confused by the terminology. The first two terms in \geqn{eq:LagrangianND} are the leading two orders in the Taylor expansion, which roots in the partial-derivative coupling term in Eq.~\eqref{eq:shiftinvariantO} that respects the axion shift symmetry. On the other hand, the third term arises due to the shift-symmetry breaking effect, as shown in Eq.~\eqref{eq:ShiftBreakO}.

Due to the presence of two different four-point vertices, the five types of diagrams in \gfig{fig:FeynmanD} can be treated as representing ten diagrams in total. 
Before we investigate the effect from the axion DM background, we first briefly review the result in the vacuum case as discussed in Ref.~\cite{Bauer:2023czj}. In the limit $r\ll 1/m_a$, the dominate contribution is from the diagram e) in \gfig{fig:FeynmanD} with the two four-point vertices being the SB type, which makes the potential $V(r)$ scales as $1/r^3$. The other nine diagrams cancel out with each other at the level of $1/r^3$ and only contributes a $1/r^5$ correction. 

However, as we are going to see below, the inclusion of axion DM background will greatly enhance the axion forces. When calculating the matrix elements of the diagrams in \gfig{fig:FeynmanD}, the axion propagators should be replaced with the background-modified one, \geqn{eq:thermalOP}.
In each diagram, when the two axion propagators both take the background term [i.e., the latter term in \geqn{eq:thermalOP}], the matrix elements vanish since the two Dirac delta functions, or equivalently the on-shell conditions, cannot be satisfied simultaneously. 
Therefore, the background contribution arises only from the cross-term, where one axion propagator is virtual while the other is on-shell from the axion DM background. Below, we show the results of $\tilde{V} (\boldsymbol{q})$ from different diagrams. The diagrams with only the SI operator involved give that
\begin{equation}
\begin{aligned}
    \hspace{-5mm}
    \tilde{V}^{SI}_a(\boldsymbol{q}) &=  \tilde{V}^{SI}_b(\boldsymbol{q})
    =
    -\frac{1}{2}
    \tilde{V}^{SI}_{c0}(\boldsymbol{q}) 
    = -\frac{1}{2} \tilde{V}^{SI}_{d0}(\boldsymbol{q}) \\
    &= \frac{1}{2}  \tilde{V}^{SI}_{e0}(\boldsymbol{q}) 
    = \frac{m_{N_1} m_{N_2} g^2_{N_1} g^2_{N_2}}{4 f^4}
    F(\boldsymbol{q})
    \label{eq:PotentialWOQ}
\end{aligned}
\end{equation}
where 
\begin{equation}
\begin{aligned}
F(\boldsymbol{q})
&\equiv
\int \frac{d^3 k}{(2 \pi)^3}
    \frac{ n(\boldsymbol{k}) }{E_k} \mathcal{A}(\boldsymbol{k},\boldsymbol{q}),
\\
     \mathcal{A}(\boldsymbol{k},\boldsymbol{q}) &\equiv
\frac{1}{-\boldsymbol{q}^2 + 2 \boldsymbol{k} \cdot \boldsymbol{q}} 
+ 
\frac{1}{-\boldsymbol{q}^2 - 2 \boldsymbol{k}\cdot \boldsymbol{q}}.
\end{aligned}
\end{equation}
We can see the explicit cancellation among these five diagrams. This is not a surprise, since the contribution from the SI term is expected to be suppressed.
Now, we consider the diagrams with the SB operator involved. We have five such diagrams in $c)$, $d)$ and $e)$ types, which contribute as
\begin{align}
     \tilde{V}^{SB}_{c1}(\boldsymbol{q}) & = 
     - \tilde{V}^{SB}_{e1}(\boldsymbol{q})
     =
    -\frac{m_{N_1} g_{N_1}^2 }{ f^2} \frac{c_{N_2}}{f^2}
    F(\boldsymbol{q}),  \nonumber\\
    \tilde{V}^{SB}_{d1}(\boldsymbol{q})& =
    - \tilde{V}^{SB}_{e2}(\boldsymbol{q})
    =
    -\frac{m_{N_2} g_{N_2}^2 }{ f^2} \frac{c_{N_1}}{f^2}
     F(\boldsymbol{q}), \nonumber\\
    \tilde{V}^{SB}_{e3}(\boldsymbol{q})& =
     2 \frac{ c_{N_1}}{f^2} \frac{ c_{N_2}}{f^2}
    F(\boldsymbol{q}).
    \label{eq:PotentialWQ}
\end{align}
$c1)$ and $d1)$ are the two triangle diagrams with the four-point vertex being the SB type. $e1)$ and $e2)$ are the bubble diagram with one vertex being the SI type and the other being the SB type.
It is straightforward to see that $c1)$ and $d1)$ cancel with $e1)$ and $e2)$. As a result, the only contribution to the potential comes from the bubble diagram with the two four-point vertices both being the SB type, $e3)$. 
A more detailed derivation of the above expressions is provided in the Supplemental Material.

%

Next, we evaluate the integral over $\boldsymbol{k}$ that occurs in the above expressions. We note that the integral is weighted by the phase space distribution $n(\boldsymbol{k})$. This means that the dominant contribution to the integral arises from $|\boldsymbol{k}|$ around 
the typical value of the momentum of the background DM, which is approximately $10^{-3}$ times of its mass and is much smaller than the nucleon mass, $|\boldsymbol{k}| \sim 10^{-3} m_a \ll m_N$. By taking this limit and also the transfer momentum $q \ll m_N$, the expression for the loop integral is greatly simplified.

Assuming an isotropic distribution $n(\boldsymbol{k})$ of DM background, we can integrate out the angular part of $\boldsymbol{k}$. Fourier transforming back to the position space, we eventually get the following expression of the background-enhanced potential $V(r)$ for the axion force,
\begin{equation}
    V(r) = 
    \frac{-  c_{N_1} c_{N_2}}{4 f^4 \pi^3 r^2}
    \int d |\boldsymbol{k}| |\boldsymbol{k}|  \frac{n(|\boldsymbol{k}|)}{E_k}
    \sin (2 |\boldsymbol{k}| r).
    \label{eq:ExpPotentialInt}
\end{equation}
To better understand the physical effects of the axion DM background on the axion force, we can further assume that DM has a top-hat distribution in a narrow momentum range $\delta |\boldsymbol{k}|$ centered around the typical DM momentum, $|\boldsymbol{k}| \sim 10^{-3} m_a$. Then, using the DM number density $n_{\rm DM}$ or the DM energy density $\rho_{\rm DM}$, we directly obtain
\begin{equation}
    n(|\boldsymbol{k}|) 
    = \frac{n_{\rm DM}}{4 \pi |\boldsymbol{k}|^2 \delta |\boldsymbol{k}| / (2 \pi)^3} 
    = \frac{2 \pi^2 \rho_{\rm DM}}{|\boldsymbol{k}|^2 \delta |\boldsymbol{k}| m_a }.
\end{equation}

Assuming the de Broglie wave length of the DM much larger than the typical distance between two nucleons, $1/|\boldsymbol{k}| \gg r$, the potential for exchanging an axion pair can be simply written as
\begin{equation}
    V(r)
    =
    \frac{- c_{N_1} c_{N_2}}{ \pi f^4}  \frac{\rho_{\rm DM}}{m^2_a} \frac{1}{r}.
    \label{eq:axionPBKG}
\end{equation}
This tells us that in the presence of the axion DM background, the scaling of the potential is enhanced to be $1/r$. Another important feature is that the magnitude of $V(r)$ is significantly enhanced by the large occupation number of axion DM particles. With the axion mass $m_a$ decreasing, such enhancement becomes increasingly pronounced.

Note that this background-enhancement effect is valid only within the range of the DM de Broglie wavelength. For distances larger than this wavelength, $r \gtrsim 1/|\boldsymbol{k}|$,  the approximation of the DM background as a coherent wave gradually breaks down. 
As a result,  the sin function in \geqn{eq:ExpPotentialInt} begins to oscillate,  leading to a damped oscillation of the background-enhanced force.

\noindent
{\bf Fifth-force constraints}--
Various experiments searching for the fifth force by measuring the attractive force between two dense objects can be applied here to constrain the background-enhanced axion force. One example is the IUPUI experiment~\cite{Chen:2014oda,Klimchitskaya:2015zpa} which aims at searching the so-called Casimir-less (CL) force. It measures the force difference between a spherical test mass and a plate made of two materials, Au and Si. The test mass is made of sapphire with radius $R = 149.3 \pm 0.2\,\mu$m; the plate has a thickness $D = 2.1\,\mu$m. For the plate placed at a distance of $\ell$ from the sphere, the force difference can be calculated as
\begin{equation}
    \begin{aligned}
    \Delta F(\ell)
    &=
    2 \pi C_s (C_{Au} - C_{Si})
    \int_{\ell}^{2 R+\ell} \mathrm{d} z_a\left[R^2-\left(z_a-R-\ell\right)^2\right]\\
    &\times
    \frac{\partial}{\partial z_a} 
    \int_{-D}^0 \mathrm{~d} z_b 
    \int_0^{\infty} r \mathrm{d} r V
    \left(\sqrt{r^2+\left(z_a-z_b\right)^2}\right)
    \end{aligned}
\end{equation}
where we used the approximation that the plate is large enough to be treated as infinite. $V(r)$ is the potential for axion force as shown in \geqn{eq:axionPBKG}. The prefactor $C_{X}$ takes different values depending on the materials of the sphere and plate,
\begin{equation}
  C_{X} \equiv A_{X} \frac{\rho_{X}}{m_{X}}  
\end{equation}
with $\rho_{X}$ the density, $m_X$ the mean mass of the atom and $A_X$ the total nucleon number of the material, respectively. The values of these parameters can be found in Ref.~\cite{1999snng.bookF}.

In addition, the background-enhanced axion force can lead to a significant violation of the 
equivalence principle (EP).
Here, we adopt the result of the Eöt-Wash experiment~\cite{Smith:1999cr} for testing EP at meter-scale distances. This experiment consists of a torsion balance with Cu 
and Pb test bodies and a 3-ton depleted uranium attractor rotating 
around it. The axion force shown in \geqn{eq:axionPBKG} would
lead to different accelerations of the test bodies made of Cu and Pb 
toward the attractor,
\begin{equation}
    \frac{\delta a}{g'} = \frac{a_{Cu} - a_{Pb}}{g'}
    = 
    \alpha \left[ 
    \left(\frac{A}{\mu}\right)_{\rm CU}
    - \left(\frac{A}{\mu}\right)_{\rm Pb}
    \right] \left(\frac{A}{\mu}\right)_{\rm U}.
\end{equation}
$\left( A/\mu \right)_i$ is the nucleon number per atomic mass unit $u$. 
$\alpha$ is defined as
\begin{equation}
 \alpha \equiv   \frac{c_N^2}{\pi f^4} \frac{\rho_{\rm DM}}{m_a^2} 
 \frac{1}{G u^2} 
 = \frac{c^4_{GG}}{f^4} \frac{4 c_1^2 m_\pi^2}{\pi} 
 \frac{\rho_{\rm DM}}{m_a^2} \frac{1}{G u^2}
\end{equation}
where $G$ is the gravitational constant. For other EP tests~\cite{Schlamminger:2007ht,Wagner:2012ui,Berge:2017ovy,Touboul:2020vkg,MICROSCOPE:2022doy} where the Earth serves as the attractor, to account for the total 
effect, we need to integrate over the volume of the entire Earth. This 
presents two key challenges. First, when the de Broglie wavelength is smaller
than the separation between the attractor and test bodies, integrating the
highly oscillatory term in calculating the background-enhanced potential is more complicated. Second, 
local topographic features around the experiment must be carefully considered. However, these issues are beyond the primary scope of this paper.
A comprehensive study of these points, as well as the possible decoherent effects 
on the background-enhanced axion force at large distances, is left for future 
work.


\begin{figure}[!t]
    \centering
    \includegraphics[width=0.486
    \textwidth]{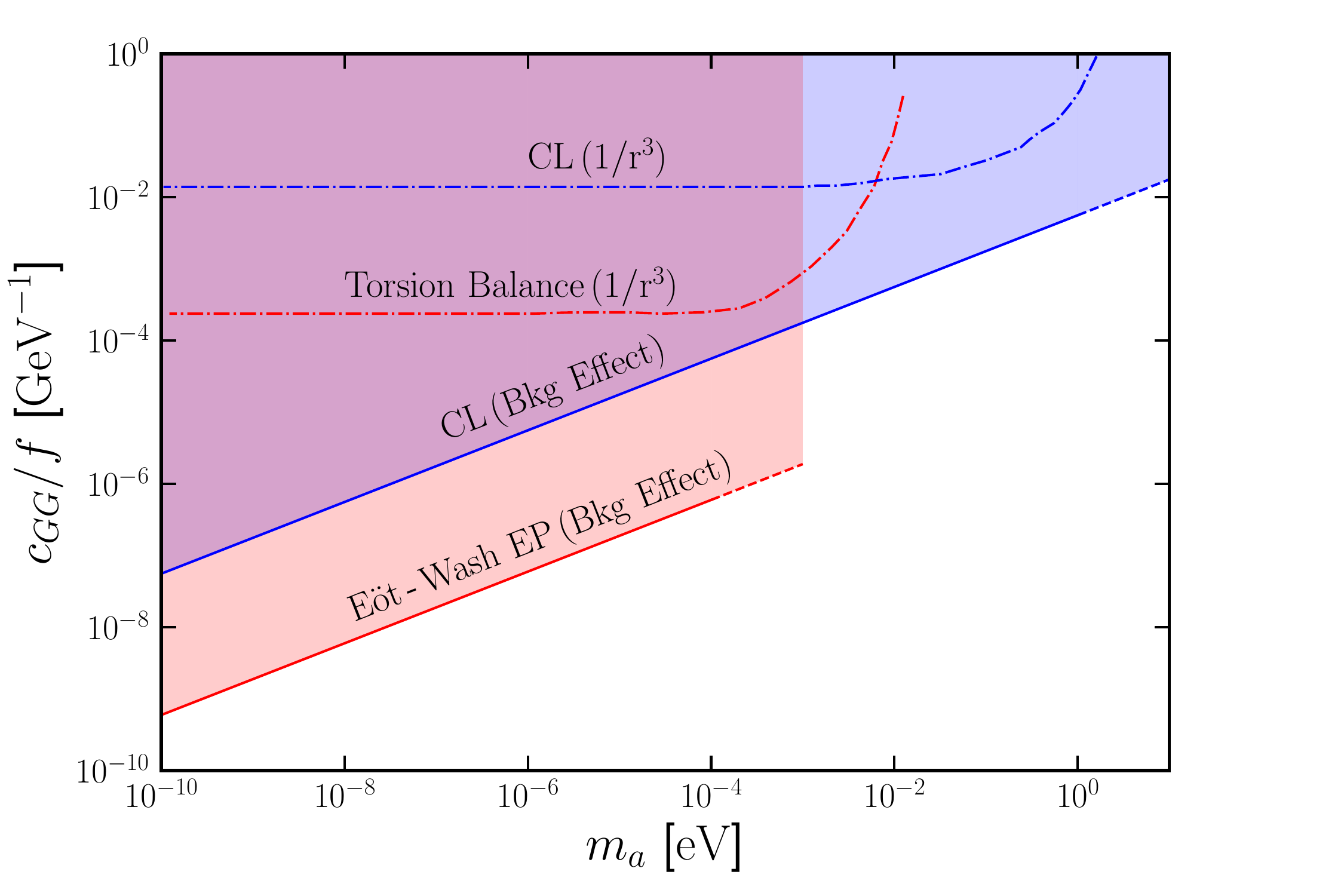}
    \caption{ The constraint on the axion parameter $c_{GG}/f$ from Casimir-less (CL) experiment~\cite{Chen:2014oda,Klimchitskaya:2015zpa} and Eöt-Wash experiment~\cite{Smith:1999cr}  are shown as blue and red solid lines, where the background effect has been considered. In comparison, we also show the constraints where only the vacuum contribution is considered~\cite{Bauer:2023czj}, denoted as the dash-dotted lines. The enhancement may not be valid in the region depicted as dashed extension lines attached to the solid lines, where the DM de Broglie wavelength starts to become shorter than the separation between the test and source masses in the experiments.}
    \label{fig:LimitAll}
\end{figure}

Based on these experiments, we can finally set constraints on the axion parameter $c_{GG}/f$, and the results are shown as solid lines in \gfig{fig:LimitAll}. 
For comparison, the constraints by considering only the vacuum contribution to the force are shown in dash-dotted lines. The blue lines are for the CL experiment, while the red lines are for the torsion-balance inverse-square law tests.
As we can see, with the background effect included, the constraints are several orders of magnitude more stringent than in the vacuum-only case. 
Additionally, as the axion mass decreases, the constraints become increasingly stringent, scaling as $m_a^{-1/2}$. The enhancement of constraints comes from two facts. First, the background effect modifies the scaling of the potential from $1/r^3$ to $1/r$. Second, the effect is proportional to the number density of axion DM and is inversely proportional to the axion mass. The dashed extension lines attached to the solid lines represent the parameter region where the DM de Broglie wavelength is comparable with the separation between the test and source masses in the two experiments. As $m_a$ increases and thus the de Broglie wavelength becomes correspondingly shorter, the constraints are expected to smoothly revert to the vacuum case.

\noindent
{\bf Discussion and conclusion}--
In this work, we have investigated the impact of the axion DM background on the spin-independent axion forces between nucleons.
This is particularly important for the axion case, as the background effect modifies only the loop-level results, and the spin-independent axion forces themselves are loop-level phenomena.
By systematically calculating the relevant Feynman diagrams, we have demonstrated that the 
axion force is significantly amplified in proportion to the number density of axion DM particles. Furthermore, the scaling of the potential for this force is modified from $1/r^3$, as in the vacuum-only case, to $1/r$ for distance $r$ smaller than the DM de Broglie wavelength.
Consequently, across a broad range of axion masses, constraints on the axion parameter $c_{GG}/f$ have been significantly enhanced by existing results from fifth-force experiments, including Casimir-less and torsion balance tests.
This underscores the fact that these types of experiments are more effective than previously believed in searching for the hypothetical particle, the axion.

An interesting future direction would involve investigating how the background effect diminishes when the typical length scale of experiments approaches or exceeds the DM de Broglie wavelength. Additionally, quantum sensors, such as atomic interferometers~\cite{Abend:2020djo,Bass:2023hoi}, atomic clocks~\cite{Peik:2020cwm,Blaum:2021eay}, or emerging experimental proposals ~\cite{Carney:2021irt,Budker:2021quh,Fierlinger:2024rdj}, may provide further insights on testing the background force of the axion.
It should be noted that for simplicity, our calculations assume an isotropic distribution of DM. A more meticulous analysis may reveal daily and/or annual modulations of the enhanced force.

~\\
{\bf Note added:} 
Upon the completion of our manuscript, we note a paper~\cite{Grossman:2025cov} on the same topic is posted on arXiv.

\section*{Acknowledgements}
We thank Bingrong Yu for his discussion on the decoherence effect and Kevin Zhou for providing his insight into a possible classical description of the background force after the first version of this paper appears.
This work is partly supported by the National Research Foundation of Korea (NRF) under Grant No. RS-2024-00405629.

\providecommand{\href}[2]{#2}\begingroup\raggedright\endgroup

\vspace{15mm}
\end{document}


\title{Background-Enhanced Axion Force by Axion Dark Matter}

\author{Yu Cheng}
\email{chengyu@kaist.ac.kr}
\affiliation{Department of Physics, Korea Advanced Institute of Science and Technology (KAIST), Daejeon 34141, South Korea}

\author{Shuailiang Ge}
\email{shuailiangge@kaist.ac.kr}
\affiliation{Department of Physics, Korea Advanced Institute of Science and Technology (KAIST), Daejeon 34141, South Korea}

\maketitle

\begin{figure}[h]
    \centering
    \includegraphics[width=0.8
    \textwidth]{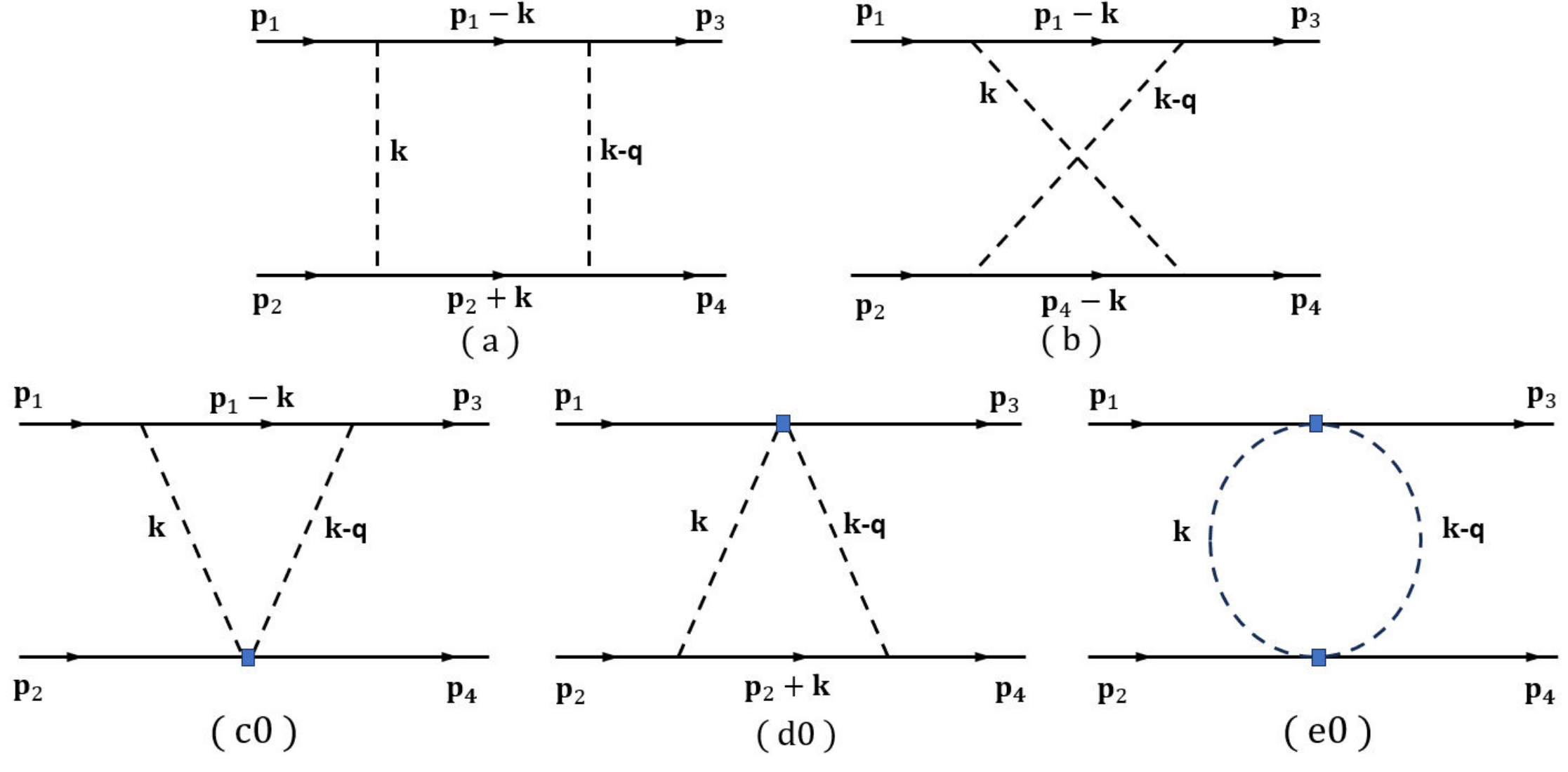}
    \caption{The Feynman diagrams generated exclusively by the shift symmetry invariant operator in the non-derivative basis. }
    \label{fig:Feyndetail}
\end{figure}

\begin{figure}[!t]
    \centering
    \includegraphics[width=0.8
    \textwidth]{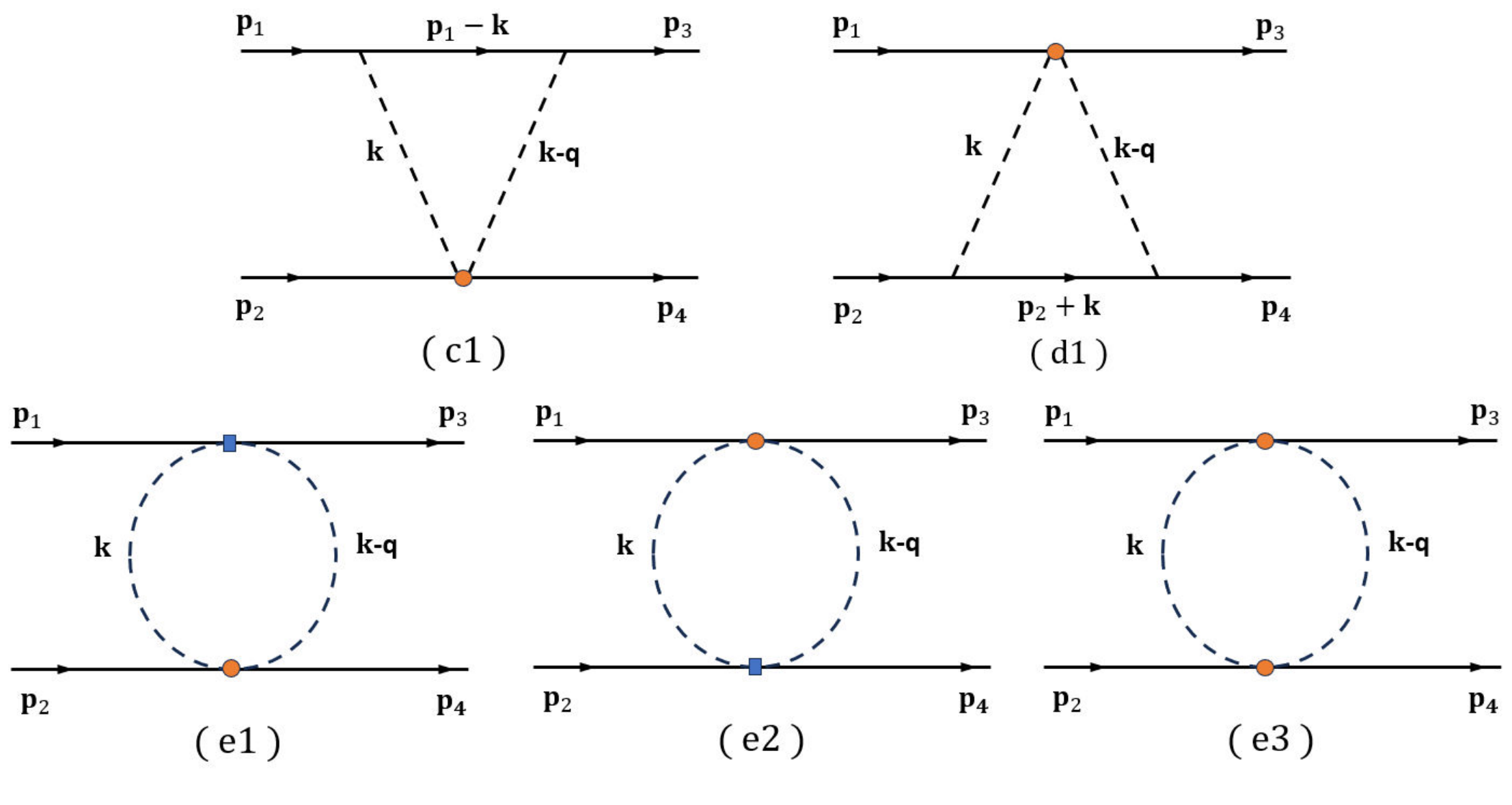}
    \caption{The additional diagrams after including the symmetry breaking operator.}
    \label{fig:Feyndetai2}
\end{figure}

In this supplemental material, we provide a detailed calculation of the spin-independent axion force in presence of the axion DM background.
A total of ten Feynman diagrams contribute to the potential, as shown in \gfig{fig:Feyndetail} and \gfig{fig:Feyndetai2}. In these diagrams, the blue box denotes the vertex arising from the operator $m_N g_N^2/(2 f^2) a^2 \bar N N $, while the orange dot represents the vertex generated by the shift symmetry breaking operator $c_N/f^2  a^2 \bar N N$.
The external momentums of the Feynman diagrams are labeled as $p_1$, $p_3$ corresponding to the initial and final states for nucleon $N_1$ and $p_2$, $p_4$ for nucleon $N_2$, respectively. The momentum for two internal lines are labeled as $k$ and $k-q$, where $q \equiv p_1 - p_3 \approx (0, \boldsymbol{q}) $ represents the momentum transfer during the scattering process.

As an example, we start with computing the potential $\tilde{V}_a$, the other nine diagrams can be obtained in the similar way. The matrix element for diagram $a)$ can be written as 
\begin{equation}
    \begin{aligned}
i \mathcal{M}_a &= 
 \frac{m^2_{N_1} g^2_{N_1}}{f^2}   \frac{ m^2_{N_2} g^2_{N_2}}{f^2}
\int \frac{d^4 k }{(2 \pi)^4}
\bar u(p_3) 
 \gamma_5 \frac{i \left( \slashed{p}_1 - \slashed{k} + m_{N_1}
\right)}{(p_1 - k)^2 - m_{N_1}^2} 
     \gamma_5 u(p_1)  \bar u(p_4) 
     \gamma_5 \frac{i \left( \slashed{p}_2 + \slashed{k} + m_{N_2}
    \right)}{(p_2 + k)^2 - m_{N_2}^2}  \gamma_5 u(p_2)\\
&\times
    \left(\frac{i}{k^2 - m_a^2} + 2 \pi n(k) \delta(k^2 - m_a^2)\right) \left(\frac{i}{(k-q)^2 - m_a^2} + 2 \pi n(k-q) \delta((k-q)^2 - m_a^2)\right)\\
    \end{aligned}
\end{equation}
keep only the crossing term which has one $\delta$ function and use the identities for fermions, $\slashed{p}_{1,2} u(p_{1,2}) = m_{N_{1,2}} u(p_{1,2})$ and $\bar u (p_{3,4}) \slashed{q} u (p_{1,2})  = 0$, we obtain
\begin{equation}
\begin{aligned}
i \mathcal{M}_a   
 &=
 \int \frac{d^4 k }{(2 \pi)^4}
\frac{m^2_{N_1} m^2_{N_2} g^2_{N_1} g^2_{N_2}}{f^4}  
\left( 2 \pi n(k) \delta(k^2 - m_a^2)\right)\\
&\times
\left[
\bar u(p_3) 
\frac{  \slashed{k} }{(p_1 - k)^2 - m_{N_1}^2} 
u(p_1) \bar u(p_4) 
\frac{ \slashed{k} }{(p_2 + k)^2 - m_{N_2}^2} 
u(p_2)  \frac{i}{(k-q)^2 - m_a^2}  \right. \\
&\left. +
\bar u(p_3) 
\frac{  \slashed{k}  }{(p_3 - k )^2 - m_{N_1}^2} 
u(p_1) \bar u(p_4) 
\frac{ \slashed{k} }{(p_4 + k )^2 - m_{N_2}^2} 
u(p_2)
\frac{i}{(k+q)^2 - m_a^2}
\right]
\end{aligned}
\end{equation}
Due to the on-shell conditions guaranteed by the $\delta$ function in the integral, 
\begin{equation}
    \delta(k^2 - m_a^2) =  \frac{1}{2 E_k} 
    \left[ \delta(k^0 - E_k) +  \delta(k^0 + E_k) \right]
    \label{eq:DeltaFRel}
\end{equation}
the denominator in the integral can be simplified to
\begin{equation}
    \begin{aligned}
    (p_{1,3} - k)^2 - m_{N_1}^2 
    &= 
   m_a^2 \mp 2 E_k m_{N_1} + 2 \boldsymbol{k}\cdot \boldsymbol{p}_{1,3} 
   \approx
   \mp 2 E_k m_{N_1}\\
    (p_{2,4} + k)^2 - m_{N_2}^2 
   &=
   m_a^2 \pm 2 E_k m_{N_2} - 2 \boldsymbol{k}\cdot \boldsymbol{p}_{2,4}
   \approx
    \pm 2 E_k m_{N_2}
    \end{aligned}
\end{equation}
where the sign in front of $E_k$ is determined by satisfying the condition in either the first or second $\delta$ function of \geqn{eq:DeltaFRel}. In the derivation, we take $|\boldsymbol{k}| \ll m_a \ll m_{N}$, as the high $\boldsymbol{k}$ contribution to the integral is exponentially suppressed by its momentum distribution. This allows the spatial component of the vector product $\boldsymbol{k} \cdot \boldsymbol{p}$ to be safely neglected. 

Recalling in the non-relativistic limit, $\bar{u}_{s^\prime} 
\left(p_3\right) \gamma^\mu u_s \left(p_1\right)
= 2 m_{N_1} \delta_0^\mu \delta_{s^\prime s}$ and 
$\bar{u}_{r^\prime} \left(p_4\right) \gamma^\mu u_r \left(p_2\right)=2 m_{N_2} \delta_0^\mu \delta_{r^\prime r}$,
again by using the identity \geqn{eq:DeltaFRel} and the approximation for the denominator in the integral, $k^0$ can be integrated out and we obtain
\begin{equation}
\begin{aligned}
    i \mathcal{M}_a
    &=
\frac{m^2_{N_1} m^2_{N_2} g^2_{N_1} g^2_{N_2}}{f^4} 
\int \frac{d^4 k}{(2 \pi)^4}
\frac{ 2 \pi n(k) }{2 E_k}
\left(
\delta(k_0 - E_k) + \delta(k_0 + E_k)
\right)
\\
&\times
\left[
2 m_{N_1} \delta_{s^\prime s} 2 m_{N_2} \delta_{r^\prime r}
\frac{  k_\mu \delta^{\mu}_0 }{\mp 2 E_k m_{N_1}} 
\frac{  k_\nu \delta^{\nu}_0 }{\pm 2 E_k m_{N_2}}
\left(
\frac{i}{k^2 + q^2 - 2 k q - m_a^2} +
\frac{i}{k^2+q^2+2kq - m_a^2}\right)
\right]\\
    &=
    -i
    \left(2 m_{N_1} \delta_{s^\prime s} 2 m_{N_2} \delta_{r^\prime r}\right)
    \frac{m_{N_1} m_{N_2} g^2_{N_1} g^2_{N_2}}{4 f^4} 
    \int \frac{d^3 k}{(2 \pi)^3}
    \frac{ n(k) }{E_k}
    \left[
    \frac{1}{-\boldsymbol{q}^2 + 2 \boldsymbol{k} \cdot \boldsymbol{q}} 
    + 
    \frac{1}{- \boldsymbol{q}^2 - 2 \boldsymbol{k}\cdot \boldsymbol{q}}
    \right]
\end{aligned}
\end{equation}
using the relation $i \mathcal{M} = - i \tilde{V} (\boldsymbol{q}) 2 m_{N_1} \delta_{s s^\prime} 2 m_{N_2} \delta_{r r^\prime}$, the potential function is determined as
\begin{equation}
    \tilde{V}_a(\boldsymbol{q}) 
    =
    \frac{m_{N_1} m_{N_2} g^2_{N_1} g^2_{N_2}}{4 f^4}
    \int \frac{d^3 k}{(2 \pi)^3}
    \frac{ n(\boldsymbol{k}) }{E_k}
    \left[
    \frac{1}{-\boldsymbol{q}^2 + 2 \boldsymbol{k} \cdot \boldsymbol{q}} 
    + 
    \frac{1}{- \boldsymbol{q}^2 - 2 \boldsymbol{k}\cdot \boldsymbol{q}}
    \right]
\end{equation}
Applying the same procedure and approximations to the other nine diagrams, it is straightforward to verify the result in Eq.~(9) and Eq.~(11) in the main text.


\bibliographystyle{utphys}
\bibliography{ref}

\vspace{15mm}